# Ideal quantum clocks and the measurement of time

Walter Gessner
Unterer Katzenbergweg 7
D-97084 Wuerzburg

## Abstract

The recently introduced concept of an "ideal quantum clock" (IQC) is extended. Especially it is shown that the time operator $T_C$ of an IQC is canonically conjugated to the Hamiltonian $H_C$ of the IQC on a certain pre-Hilbert space. Further it is discussed how the IQC interacts with another physical system D and to what extent the IQC measures the time differences of any prescribed initial and final states of D.

## 1. Ideal quantum clocks and operator time.

The concept of an IQC is based on three assumptions [ 1 ]:

1. An IQC C is a closed physical system with states $\varphi_C$ in a pre-Hilbert-space. The time evolution $\varphi_C(t) := U(t)\varphi_C$ of any $\varphi_C$ satisfies Schrödinger's equation with a Hamiltonian $H_C$. Thus $\varphi_C(t)$ is called Schrödinger curve.
2. Any ideal clock has a certain time resolution $\tau > 0$ so that the clock provides (by the "clicks") equidistant time points $\{0, \pm\tau, \pm 2\tau, ...\}$ which exclude one another.
3. Any ideal clock can be setted in such a way that the clock delivers the time points $\{t, t\pm\tau, t\pm 2\tau, ...\}$ instead of $\{0, \pm\tau, \pm 2\tau, ...\}$ where t is any real parameter. The state space to be defined has to be independent of this parameter (time invariance of the state space).

Translating these properties into the traditional quantum theoretical formalism one gets the definition [ 1 ] of an IQC, where $\varphi_C(t)$ is the normed Schrödinger curve of 1 and $U(t) := \exp(-iH_C)$:

$$<\varphi_C(\tau^m)|\varphi_C(\tau^n)> = \delta^{mn} \text{ for all } m, n \in \mathbb{Z}, \text{ where } \tau^n := n\tau, n \in \mathbb{Z}. \tag{01}$$

$$U(t)S_C = S_C \text{ for all t, where } S_C := \text{span}\{\varphi_C(\tau^n)|n \in \mathbb{Z}\}. \tag{02}$$

For reasons of convergence of the time operator defined below, the subspace $s_C$ of $S_C$ of all $\varphi := \sum_{n=-\infty}^{+\infty} d^n \varphi_C(\tau^n)$ with complex valued $d^n$ satisfying $\sum_{n=-\infty}^{+\infty} |n||d^n| < \infty$ is to be introduced. $s_C$ is pre-Hilbert space [ 2 ] and dense in $S_C$. The space $s_C$ is time invariant also: $U(t)s_C = s_C$.

The definition of the time operator needs the mapping $P_C: s_C \rightarrow S_C$ which orders to any $\varphi := \sum_{n=-\infty}^{+\infty} d^n \varphi_C(\tau^n) \in s_C$ all those $\tau^n$ with corresponding probabilities whose $\varphi_C(\tau^n)$ arise in the expansion of $\varphi$:

$$P_C \varphi := \sum_{n=-\infty}^{+\infty} d^n \tau^n \varphi_C(\tau^n). \tag{03}$$

The time operator $T_C: s_C \to S_C$ of C is then

$$T_C := \tau^{-1} \int_{-\tau/2}^{+\tau/2} du\, U^+(u) P_C U(u). \tag{04}$$

In [ 1 ] the main features of $T_C$ are proven. The decisive result reads

$\langle\varphi|[T_C, H_C]|\varphi\rangle = i$ for all $\varphi \in s_C$ with $\|\varphi\| = 1$. (05)

From this follows immediately the time-energy uncertainty relation $\sigma(T_C)\sigma(H_C) \geq \frac{1}{2}$ with respect to the isolated IQC, where the standard deviation $\sigma(T_C)$, and accordingly $\sigma(H_C)$, is

$\sigma(T_C)^2 = \text{Var}(T_C) = \langle T_C\varphi|T_C\varphi\rangle - \langle\varphi|T_C|\varphi\rangle^2.$ (06)

The result $\langle\varphi|[T_C, H_C]|\varphi\rangle = i$ can now be extended in the following way:

**Theorem:**

$T_C$ and $H_C$ are canonically conjugated on the pre-Hilbert space $s_C$:

$[T_C, H_C] = i$, in the integral form $(t \in \mathbb{R},\, U(t) = \exp(-iH_C))$ (07)

$[T_C, U(t)] = tU(t).$ (08)

**Proof:**

In [ 1 ] it is proven that $\langle\varphi|[T_C, H_C]|\varphi\rangle = i$ for all $\varphi \in s_C$ with $\|\varphi\| = 1$. Therefore, the symmetric

$K := i[T_C, H_C] + 1$ satisfies $\langle\varphi|K|\varphi\rangle = 0$ for all $\varphi \in s_C$. (09)

Assume now that $\varphi, \psi \in s_C$ do exist so that $\langle\varphi|K|\psi\rangle \neq 0$. Then $\langle\varphi|K|\psi\rangle$ can be written as $\langle\varphi|K|\psi\rangle = e^{i\alpha}|\langle\varphi|K|\psi\rangle|$ with $\alpha \in [0, 2\pi[$.

The element $\varphi + e^{i\beta}\psi \in s_C$, where $\beta$ is any parameter, satisfies now

$0 = \langle\varphi + e^{i\beta}\psi|K|\varphi + e^{i\beta}\psi\rangle = e^{-i\beta}\langle\psi|K|\varphi\rangle + e^{+i\beta}\langle\varphi|K|\psi\rangle =$ (10)

$e^{+i(\alpha+\beta)}|\langle\varphi|K|\psi\rangle| + e^{-i(\alpha+\beta)}|\langle\varphi|K|\psi\rangle| = 2\cos(\alpha+\beta)|\langle\varphi|K|\psi\rangle|$

so that $\cos(\alpha+\beta) = 0$.

The choice $\beta := -\alpha$ for example leads to a contradiction. Therefore

$\langle\varphi|K|\psi\rangle = 0$ for all $\varphi, \psi \in s_C$. (11)

Let now be $K\varphi = \sum_{n=-\infty}^{+\infty} d^n \varphi_C(\tau^n) \in S_C$ with $\varphi \in s_C$. Because $\varphi_C(\tau^m) \in s_C$, (11) yields $0 = \langle\varphi_C(\tau^m)|K|\varphi\rangle = d^m$ for all m so that $K\varphi = 0$ for all $\varphi \in s_C$. Summarizing $[T_C, H_C] = i$ on $s_C$.

$U(t) = \exp(-itH_C)$ yields

$U^+(t)T_CU(t) = T_C + it[H_C, T_C] = T_C + t$, so that

$T_CU(t) = U(t)T_C + tU(t)$. This is the assertion. ∎

As an application, the time invariance of the standard deviation $\sigma(T_C)$ is proven:

$\langle T_C\varphi(t)|T_C\varphi(t)\rangle - \langle\varphi(t)|T_C|\varphi(t)\rangle^2 =$
$\langle T_CU(t)\varphi|T_CU(t)\varphi\rangle - \langle\varphi|U^+(t)T_CU(t)|\varphi\rangle^2 =$
$\langle U^+(t)T_CU(t)\varphi|U^+(t)T_CU(t)\varphi\rangle - \langle\varphi|U^+(t)T_CU(t)|\varphi\rangle^2 =$
$\langle (T_C + t)\varphi|(T_C + t)\varphi\rangle - \langle\varphi|(T_C + t)\varphi\rangle^2 = \langle T_C\varphi|T_C\varphi\rangle - \langle\varphi|T_C|\varphi\rangle^2$,

so that $\sigma(T_C)$ is independent of t.

## 2. The non-existence of eigenstates of $T_C$ and $H_C$ in $s_C$.

The concept of an IQC forbids the existence of eigenstates of $H_C$ and of $T_C$ in $s_C$ because $\langle\varphi|[T_C, H_C]|\varphi\rangle = 0$, if $\varphi \in s_C$ is such an eigenstate, whereas $\langle\varphi|[T_C, H_C]|\varphi\rangle = i$, $\|\varphi\| = 1$. This non-existence of eigenstates is not only a formal result but has physical backgrounds:

Eigenstates of $H_C$ cannot "age" so that they don't carry any time information: Let namely $\varphi \in s_C$, $\|\varphi\| = 1$, be any eigenstate of $H_C$ with the eigenvalue h so that $H_C\varphi = h\varphi$ and $\varphi(t) := U(t)\varphi = \exp(-iht)\varphi$. Then for all t

$\langle\varphi(t)|T_C|\varphi(t)\rangle = \langle\varphi(0)|T_C|\varphi(0)\rangle$.

Eigenstates of $T_C$ don't allow the time evolution at all.
Proof: Let $T_C\varphi = t_\varphi\varphi$ with $t_\varphi \in \mathbb{R}$, $\varphi \in s_C$ and $\|\varphi\| = 1$. One gets first

$U(t)T_C\varphi = t_\varphi U(t)\varphi$, furtheron $(T_C - t)U(t)\varphi = t_\varphi U(t)\varphi$ and

$T_C\varphi(t) = (t + t_\varphi)\varphi(t)$ where $\varphi(t) := U(t)\varphi$. In this way, $\varphi(t)$ is an eigenstate of $T_C$ to any t. Because of the symmetric $T_C$, eigenstates with different eigenvalues are orthogonal. Therefore

$\|\varphi(t+dt) - \varphi(t)\|^2 = \langle\varphi(t+dt) - \varphi(t)|\varphi(t+dt) - \varphi(t)\rangle = +2$ for any $dt \neq 0$ so that the derivation $\lim[\varphi(t+dt) - \varphi(t)](dt)^{-1}$ for $dt \to 0$ does not exist. As a conse-quence, φ(t) cannot satisfy Schrödinger's equation Hφ(t) = i d/dt φ(t), so that the time evolution of φ is impossible.

## 3. The time measurement by an IQC.

Let now a closed physical system D with a Hamiltonian $H_D$ be given. It is assumed that D principally can be described by a Schrödinger curve ψ(t), t∈ℝ, in a pre-Hilbert state space $S_D$. The question is to what extent the IQC measures the time difference of any prescribed initial and final states of D.

First of all, both systems C and D are to be combined to a system C+D according to possible interactions between C and D. Thus it were convenient to get a state space $S_{CD}$ which contains $S_C$ (and so $s_C$) and $S_D$ as subspaces and to have $H_{CD}:= H_C + H_D$ as the Hamiltonian of C+D. But $S_C$ and $s_C$ admit $H_C$ as the only Hamiltonian because of their required time invariance (02). Therefore a condition of compatibility is to be introduced:

**Definition:**
The systems C and D are **compatible**, if the restriction of the Hamiltonian $H_D$ to the spaces $S_C$ resp. $s_C$ is $H_C$.

The existence of $S_{CD}$ and this compatibility provided, the time evolution of C+D may then given by the Schrödinger curve $φ_{CD}(t):= φ_C(t)+ψ(t)∈S_{CD}$ defined by a Hamiltonian $H_{CD}$ usually assumed as $H_C + H_D$.
Interactions between C and D arise if ψ(t) has a nonvanishing projection into $s_C$. Then C is disturbed by D, and D is modified by C. The following cases are possible:

### 3.1  No interaction between C and D:

Then $S_C \cap S_D = \{0\}$, and a suitable state space of C+D is the direct sum $s_C \oplus S_D$. Both systems run independently (and are compatible), and the Schrödinger curve is $φ_C(t) \oplus ψ(t)$, with $\|φ_C(t)\| = 1$ and $ψ(t) \in S_D$. The Hamiltonians $H_C$, $H_D$ operate separately on $s_C$ resp. $S_D$. The set of all these states is $\{φ_C(t) \oplus ψ(t) | t \in \mathbb{R}\}$, where the pairs $φ_C(t) \oplus ψ(t)$ are "coupled" by the same value of t. Let now be given any two states $\varphi_C^1 \oplus \psi^1$ and $\varphi_C^2 \oplus \psi^2$ from this set. $\psi^1$ may be an initial state of the process D, $\psi^2$ a final state. Then, the coupled $\varphi_C^1$, $\varphi_C^2$ yield the expectation values $<\varphi_C^i|T_C|\varphi_C^i>$, i = 1, 2, of the time operator defined by the IQC. These $<\varphi_C^i|T_C|\varphi_C^i>$ are to be ordered to the corresponding $\psi^i$. The difference $|<\varphi_C^2|T_C|\varphi_C^2> - <\varphi_C^1|T_C|\varphi_C^1>|$ is independent of the chosen zero-point of the time given by $\varphi_C(0)$ and is taken as the duration of the process in question delivered by the IQC. Because of the coupling $\varphi_C(t) \oplus \psi(t)$, the relation $<\varphi_C(t)|T_C|\varphi_C(t)> = t$ [ 1 ], equ. (41), yields then t as the expectation value of the time operator also with respect to $\psi(t)$. Summarizing, the time measurement of D plays completely in $s_C$ and all features of an isolated C, especially the time-energy uncertainty relation, can be applied to C+D also.

**3.2 Weak interaction between C and D:**

It is now assumed that a state space $S_{CD}$ exists which contains $S_C$ (and so $s_C$) and $S_D$ as subspaces. The Schrödinger curve $\varphi_{CD}(t) := \varphi_C(t)+\psi(t) \in S_{CD}$ describes the system C+D.

Let now C be given by $\varphi_C(t) = \sum_{n=-\infty}^{+\infty} c^n(t)\varphi_C(\tau^n) \in s_C$ and the projection of $\psi(t)$ into $S_C$ by $\psi_C(t) := \sum_{n=-\infty}^{+\infty} d^n(t)\varphi_C(\tau^n)$. It is assumed that $\psi_C(t) \in s_C$ (the case $\psi_C(t) \in S_C$ but $\psi_C(t) \notin s_C$ is mentioned in 3.3). The compatibility above yields $\psi_C(t) = \exp(-itH_C)\psi_C(0)$. The IQC C is now disturbed by $\psi_C(t)$ and has changed in the non-ideal quantum clock C($\psi$) with the Schrödinger curve $\varphi_C(t)+\psi_C(t) \in s_C$. If the interaction between C and D is weak enough, the expectation values of the corresponding time operator $T_{C(\psi)}$ for any prescribed initial and final states $\varphi_C^i + \psi_C^i$, i = 1, 2, from the set $\{\varphi_C(t)+\psi_C(t)|t \in \mathbb{R}\}$ are to be defined as the time values of the corresponding $\psi^i$ given by C($\psi$). All this runs in the same way as discussed in 3.1. But C($\psi$) approximates only the properties of C.

The interaction between C and D is connected to an exchange of energy given for example by the difference of the expectation values of $H_C$ with respect to $\varphi_C(0)$ resp. $\varphi_C(0)+\psi_C(0)$.

### 3.3 Strong interaction between C and D:

In 3.2 the case $\psi_C(t) \notin s_C$ was excluded. Let now $\psi_C(t) \in S_C$ but $\psi_C(t) \notin s_C$ for certain t. Then $\varphi_C(t)+\psi_C(t)$ are in general outside of the domain of $T_C$. In this case C is disturbed too much so that C cannot work. A non-ideal, but still usable, quantum clock $C(\psi)$ as above does not exist.

## 4. A remark about the "theorem of Pauli".

The argument of Pauli against any (selfadjoint) time operator T exploits the properties of the formal unitary operator $\exp(+ikT)$ with any real parameter k, especially the equation $\exp(-ikT)H\exp(+ikT) = H + k$, which follows from the assumed identity $[T, H] = i$.

Accordingly, in the IQC-theory the operator

$$V(k) := \exp(+ikT_C): s_C \to S_C \tag{12}$$

to any fixed real k is to be discussed. First of all, as a consequence of the domain $s_C$ of $T_C$, the terms $T_C^2$, $T_C^3$ and so on, after all V(k), are not defined on $s_C$ in general, so that the maximal domain $D_{V(k)} \subseteq s_C$ of V(k) may be the trivial subspace $\{0\}$ of $s_C$ for certain k. At least the common domain of all V(k) is the trivial subspace of $s_C$:

$$D_V = \{0\}, \text{ where } D_V := \bigcap D_{V(k)} \text{ for all k.} \tag{13}$$

Proof:
Let be $\varphi \in D_V$, $\|\varphi\| = 1$ and $\varphi_k := V(k)\varphi$. Then $\varphi_k \in S_C$ for all k. In [ 1 ], Lemma 1 c, it is proven that a maximal energy W exists so that all expectation values $<\varphi|H_C|\varphi>$ with $\varphi \in S_C$ and $\|\varphi\| = 1$ satisfy

$$|<\varphi|H_C|\varphi>| \leq W. \tag{14}$$

As elements of $S_C$, $\varphi$ as well as $\varphi_k$ satisfy
$|<\varphi|H_C|\varphi>| \leq W$ and $|<\varphi_k|H_C|\varphi_k>| \leq W$. On the other hand
$<\varphi_k|H_C|\varphi_k> = <\varphi|V(-k)H_CV(+k)|\varphi> = <\varphi|H_C|\varphi> + k$
because of $\exp(-ikT_C)H_C\exp(+ikT_C) = H_C + [-ikT_C, H_C]$, so that

$$|<\varphi|H_C|\varphi> + k| = |<\varphi_k|H_C|\varphi_k>| \leq W \text{ for all k.} \tag{15}$$

Contradiction. Therefore, $\varphi \in D_V$ with $\|\varphi\| = 1$ cannot exist, so that $\varphi = 0$. ∎